\documentclass[aps,prd]{revtex4-2}

\usepackage{amsmath,amssymb}
\usepackage{graphicx}
\usepackage{bm}
\usepackage{array}
\newcolumntype{P}[1]{>{\centering\arraybackslash}p{#1}}

\begin{document}

\title{Self-Consistent Determination of the Transition Temperature Between the
$^{14}\mathrm{C}(n,\gamma)^{15}\mathrm{C}$ and
$^{14}\mathrm{C}(p,\gamma)^{15}\mathrm{N}$ Reactions
 }

\author{R. Ya. Kezerashvili$^{1,2,3}$, N. A. Burkova$^{4}$, A. S. Tkachenko$^{5}$, S. B. Dubovichenko$^{5}$}
\affiliation{
$^{1}$New York City College of Technology, The City University of New York,
Brooklyn, NY, USA\\
$^{2}$The Graduate School and University Center, The City University of New
York, New York, NY, USA\\
$^{3}$Long Island University, Brooklyn, NY, USA\\
$^{4}$al-Farabi Kazakh National University, Almaty, Kazakhstan \\
$^{5}$Department of Science, META University, Almaty, Kazakhstan \\
}




\begin{abstract}
We present the first self-consistent theoretical study of the competing
$^{14}\mathrm{C}(n,\gamma)^{15}\mathrm{C}$
and
$^{14}\mathrm{C}(p,\gamma)^{15}\mathrm{N}$
reactions within the same modified potential cluster model (MPCM). For the
$^{14}$C$(p,\gamma_{0})^{15}$N reaction,
total cross sections, astrophysical $S$ factors, and reaction rates are calculated using interaction potentials constrained by the available scattering and bound-state data.
The astrophysical $S$-factor is estimated as $S(0)=4.5(1)$~keV$\cdot \text{b}$. 
Combining these results with our recent MPCM calculations for
$^{14}\mathrm{C}(n,\gamma)^{15}\mathrm{C}$,
we determine the transition temperature at which proton capture overtakes neutron capture in the production of
$^{15}\mathrm{N}$. The self-consistent comparison predicts a transition temperature
$T_9^{\rm c.p.}=2.5$
under Maxwell--Boltzmann statistics, significantly higher than previous estimates. 
The analysis is extended to Tsallis statistics, demonstrating that deviations from thermal equilibrium produce substantial shifts of the transition temperature. These results provide improved nuclear-physics input for astrophysical nucleosynthesis calculations.

\end{abstract}
\date{\today }
\maketitle


\textbf{\textit{{Introduction.} }} 
Radiative neutron- and proton-capture reactions play an important role in stellar nucleosynthesis by governing the competition between reaction pathways leading to the synthesis of heavier nuclei.
The competition between the
$^{14}$C$(n,\gamma)^{15}$C and
$^{14}$C$(p,\gamma)^{15}$N reactions determines the dominant pathway for the production of
$^{15}$N under different astrophysical conditions.
We further examine how this transition is modified within the framework of non-extensive Tsallis statistics.
Such competition may become particularly significant in neutron-rich astrophysical environments, such as carbon-rich regions of asymptotic giant branch (AGB) stars and intermediate neutron-capture (i-process) sites, where neutron densities and departures from thermal equilibrium can substantially modify nucleosynthesis pathways ~\cite{Ref1,Ref2,Ref3,Ref4,Ref6,Ref7}. 

The long-lived isotope $^{14}\mathrm{C}$ provides a unique branching point because both reactions
$^{14}\mathrm{C}(n,\gamma)^{15}\mathrm{C}
(\beta^-)
^{15}\mathrm{N}$
and
$^{14}\mathrm{C}(p,\gamma)^{15}\mathrm{N}$
lead to the production of the stable isotope $^{15}\mathrm{N}$. While the neutron-capture reaction has been investigated experimentally \cite{Ref19,Ref20,Ref21,Ref22,Ref23} and theoretically ~\cite{Ref24,Ref25,Ref26,Ref27,Ref28,Ref49}, during the last three decades, 
experimental information for the
$^{14}\mathrm{C}(p,\gamma)^{15}\mathrm{N}$ reaction remains limited to a single set of low-energy measurements~\cite{Ref29}. Consequently, reliable theoretical calculations of the proton-capture cross section and reaction rate are essential for astrophysical applications~\cite{Ref30,Ref31}.

Previous comparisons of the
$^{14}\mathrm{C}(n,\gamma)^{15}\mathrm{C}$ and
$^{14}\mathrm{C}(p,\gamma)^{15}\mathrm{N}$ reaction rates were based on calculations performed within different theoretical approaches~\cite{Ref32,Ref33,Ref34}. Such heterogeneous comparisons introduce systematic uncertainties into the determination of the temperature at which proton capture becomes more efficient than neutron capture. A self-consistent comparison of both competing reactions based on a common theoretical framework has not been available.

In the present work, both radiative-capture reactions are investigated within the modified potential cluster model
(MPCM). We combine our recent MPCM calculation of the
$^{14}$C$(n,\gamma)^{15}$C reaction~\cite{Ref24} with a new analysis of the
$^{14}$C$(p,\gamma)^{15}$N process within the same theoretical framework.
This enables the first self-consistent determination of the transition temperature between neutron- and proton-capture dominance under both Maxwell--Boltzmann and Tsallis statistics.
The proton-capture cross section, astrophysical $S$-factor, and thermonuclear reaction rate are calculated to establish this comparison.

\textbf{\textit{Theoretical Framework.}} 
The radiative capture reactions
$^{14}\mathrm{C}(n,\gamma)^{15}\mathrm{C}$ and
$^{14}\mathrm{C}(p,\gamma)^{15}\mathrm{N}$
are investigated within the framework of the MPCM, in which the interacting nuclei are treated as two-body cluster systems and the intercluster interaction is described by phenomenological potentials whose orbital states are classified according to Young diagrams. The interaction potentials are constrained by elastic-scattering phase shifts, resonance energies and widths, binding energies, asymptotic normalization coefficients, and other spectroscopic characteristics of the corresponding bound states. Radiative-capture cross sections are calculated from the electric dipole $E1$-transition matrix elements between the scattering and bound-state wave functions. Details of the MPCM formalism, construction of the interaction potentials, and calculation of radiative-capture cross sections are given in Refs.~\cite{Ref35,Ref62,Ref24,RefXX,RefYY}.

The thermonuclear reaction rate is obtained by averaging the calculated capture cross section over the Maxwell--Boltzmann energy distribution~\cite{Ref3} 
\begin{equation}
N_A\langle\sigma v\rangle =
3.7313\times10^{4}\,
\mu^{-1/2}
T_9^{-3/2}
\int_{0}^{\infty}
\sigma(E)\,
E\,
\exp\!\left(-11.605\,\frac{E}{T_9}\right)\,
dE,
\label{Rate_MB}
\end{equation}
where $N_A$ is Avogadro number, $\mu$ is the reduced mass in a.m.u., $\sigma(E)$ is the capture cross section in $\mu$b, $E$ is the center-of-mass energy in MeV, and the reaction rate
$N_A\langle\sigma v\rangle$
is expressed in units of
$\mathrm{cm^3\,mol^{-1}\,s^{-1}}$.

To investigate possible deviations from thermal equilibrium, following the formalism developed by Hou \textit{et al.}~\cite{Ref66}, the reaction rates are additionally evaluated within the non-extensive Tsallis statistical framework,

\begin{equation}
N_A\langle\sigma v\rangle_q
=
B_q
\sqrt{\frac{8}{\pi\mu}}
\,
\frac{N_A}
{(k_BT)^{3/2}}
\int_0^{E_{\rm max}}
\sigma(E)\,
E
\left[
1-(q-1)\frac{E}{k_BT}
\right]^{\frac{1}{q-1}}
dE,
\label{Rate_Tsallis}
\end{equation}
where $q$ is the non-extensive parameter, $B_q$ is the normalization constant, and
%
$E_{\rm max}
=
\frac{k_BT}{q-1}$, $k_BT
=
0.086173\,T_9
\ {\rm MeV},$
%
for $q>1$, while $E_{\rm max}\rightarrow\infty$ for $q<1$. 
The reaction rates are calculated over the temperature interval
$0.01\le T_9\le10$.


\textbf{\textit{Results and Discussion.}} The neutron-capture reaction
$^{14}\mathrm{C}(n,\gamma)^{15}\mathrm{C}$
has recently been investigated in ~\cite{Ref24} within the same MPCM framework employed in the present work for the
$^{14}\mathrm{C}(p,\gamma)^{15}\mathrm{N}$
reaction. The good agreement between the MPCM calculations~\cite{Ref24} and the first \emph{ab initio} prediction of Navr\'atil \textit{et al.}~\cite{Ref26} for the cross sections, obtained within an entirely different theoretical framework, provides particularly strong evidence for the reliability of the calculated neutron-capture reaction rate. Consequently, both competing reaction rates considered in the present work are determined within a common theoretical framework, eliminating systematic uncertainties associated with combining results from different models and enabling a fully self-consistent comparison of the
$^{14}\mathrm{C}(n,\gamma)^{15}\mathrm{C}$
and
$^{14}\mathrm{C}(p,\gamma)^{15}\mathrm{N}$
reactions.

\begin{figure*}[t]
\centering
\begin{minipage}{0.48\textwidth}
    \centering
    \includegraphics[width=\textwidth]
    {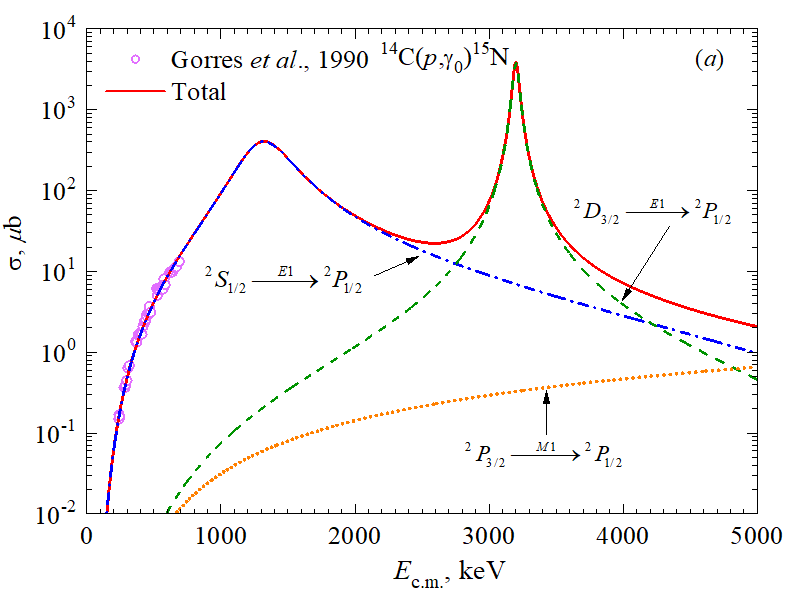}
\end{minipage}
\hfill
\begin{minipage}{0.48\textwidth}
    \centering
    \includegraphics[width=\textwidth]
    {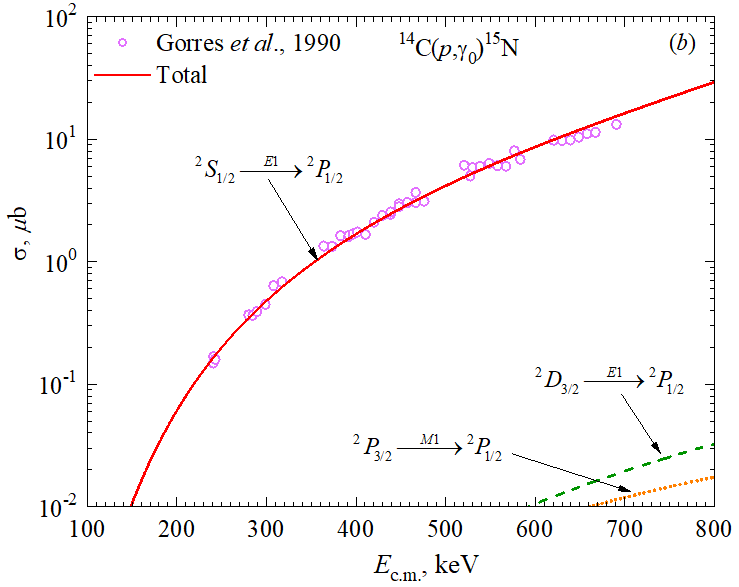}
\end{minipage}
\caption{Total and partial cross sections of the proton radiative capture
$^{14}\mathrm{C}(p,\gamma_{0})^{15}\mathrm{N}$.  The total cross section (solid red curve) is the sum of the three partial transitions to the ground state of
$^{15}\mathrm{N}$,
$
\sigma_{\rm tot}
=
\sigma\!\left(
{}^{2}S_{1/2}\rightarrow{}^{2}P_{1/2}
\right)
+
\sigma\!\left(
{}^{2}D_{3/2}\rightarrow{}^{2}P_{1/2}
\right)
+
\sigma\!\left(
{}^{2}P_{3/2}\rightarrow{}^{2}P_{1/2}
\right)$. MPCM calculations are shown for:
$(a)$ $E_{\mathrm{c.m.}}=140$--$5000$~keV and
$(b)$ $E_{\mathrm{c.m.}}=140$--$800$~keV.
The partial and total cross sections are indicated in the figure panels. Experimental data are
from Ref.~\cite{Ref29}.}
\label{fig:pgamma_cross_sections}
\end{figure*}

Having established the reliability of the neutron-capture reaction rate, we now consider the proton-capture reaction. The proton-capture reaction
$^{14}\mathrm{C}(p,\gamma)^{15}\mathrm{N}$
was investigated within the MPCM  using interaction potentials constrained by the available scattering and bound-state properties. The calculated total and partial radiative-capture cross sections for the
$^{14}$C$(p,\gamma)^{15}$N
reaction are shown in Fig.~\ref{fig:pgamma_cross_sections}.
The calculation reproduces the measured low-energy cross sections over the entire experimental energy interval without introducing additional parameters in the capture calculation. The observed resonant structure of the total cross section is produced by direct $E1$  transitions from the experimentally established
$J^\pi=1/2^+$ and $3/2^+$ resonance scattering states to the ground state of
$^{15}$N.
At low energies, the enhancement of the cross section is governed primarily by the low-energy tail of the
$J^\pi=1/2^+$ resonance.

\begin{figure*}[t]
\centering
\begin{minipage}{0.48\textwidth}
    \centering
    \includegraphics[width=\textwidth]{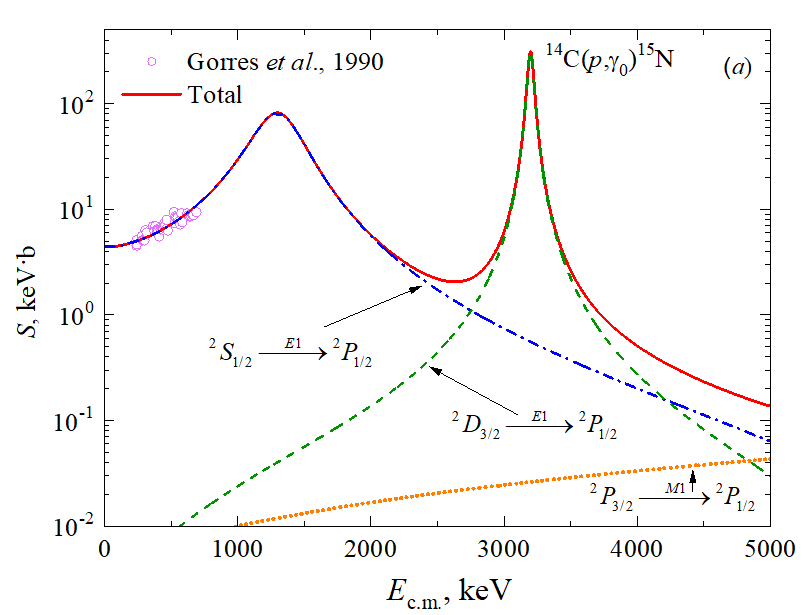}
\end{minipage}
\hfill
\begin{minipage}{0.48\textwidth}
    \centering
    \includegraphics[width=\textwidth]{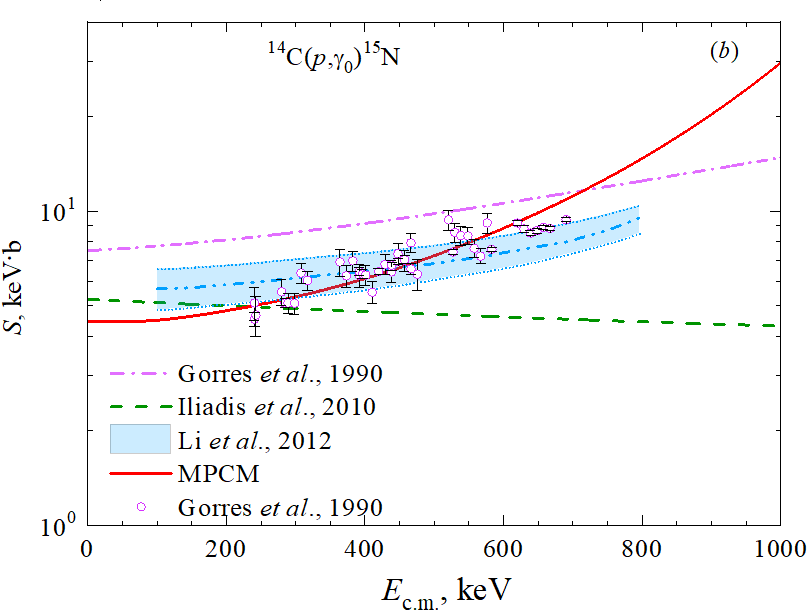}
\end{minipage}
\caption{Astrophysical $S$-factor of the proton radiative capture
$^{14}\mathrm{C}(p,\gamma_{0})^{15}\mathrm{N}$.  The solid curves represent the present MPCM calculations. ($a$)
Notations for the partial contributions are the same as in Fig.~\ref{fig:pgamma_cross_sections}. $(b)$ Comparison of the present MPCM astrophysical $S$-factor with previous calculations \cite{Ref29,Ref30,Ref61} and the experimental data \cite{Ref29}. }
\label{fig:Sfactor}
\end{figure*}

The corresponding astrophysical $S$-factor is presented in Fig.~\ref{fig:Sfactor}. The calculated curve reproduces the available experimental data and exhibits the expected smooth energy dependence below the Coulomb barrier. The astrophysical $S$-factor averaged over the low-energy interval $E_{c.m.} \sim 100--140$ keV is near constant within the deviation of $\sim$ 2\%. Extrapolation toward zero energy yields
$
S(0)=4.5(1)\ \rm keV\cdot{b}.
$

\begin{figure}[t]
\centering
\includegraphics[width=0.45\linewidth]{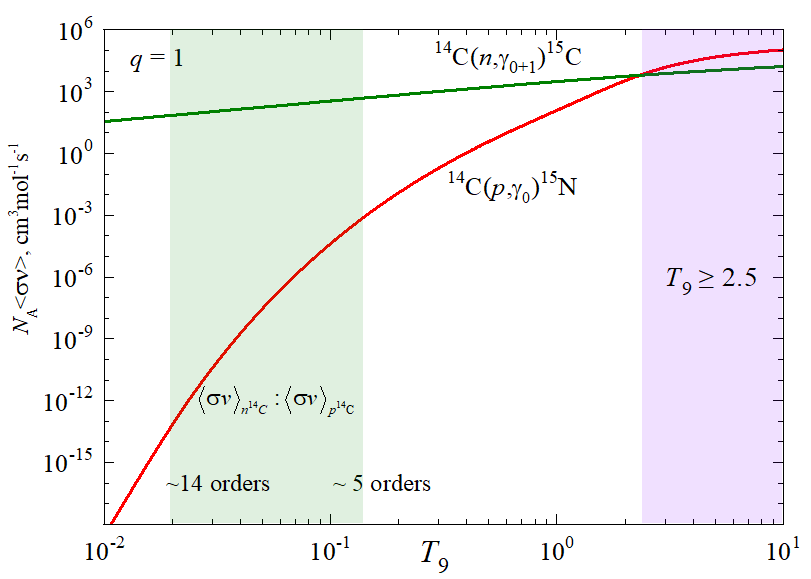}
\caption{
Comparison of the Maxwell--Boltzmann thermonuclear reaction rates calculated
within the MPCM for the
$^{14}\mathrm{C}(n,\gamma_{0+1})^{15}\mathrm{C}$ and
$^{14}\mathrm{C}(p,\gamma_{0})^{15}\mathrm{N}$ reactions.
The green shaded region ($T_9\simeq0.02$--0.15) corresponds to temperatures
typical of hydrogen burning in AGB stars.
The reaction rate of
$^{14}\mathrm{C}(n,\gamma_{0+1})^{15}\mathrm{C}$
exceeds that of
$^{14}\mathrm{C}(p,\gamma_{0})^{15}\mathrm{N}$
throughout this temperature range.
The purple shaded region ($T_9\gtrsim2.5$) marks the temperatures at which the
proton-capture reaction becomes dominant.
The intersection of the two curves at
$T_9^{\mathrm{c.p.}}\approx2.5$
defines the transition from neutron-capture to proton-capture dominance in the
production of $^{15}\mathrm{N}$.
}
\label{fig:ReactionRatesComparison}
\end{figure}
%
\begin{figure}[t]
\centering
\includegraphics[width=0.4\linewidth]{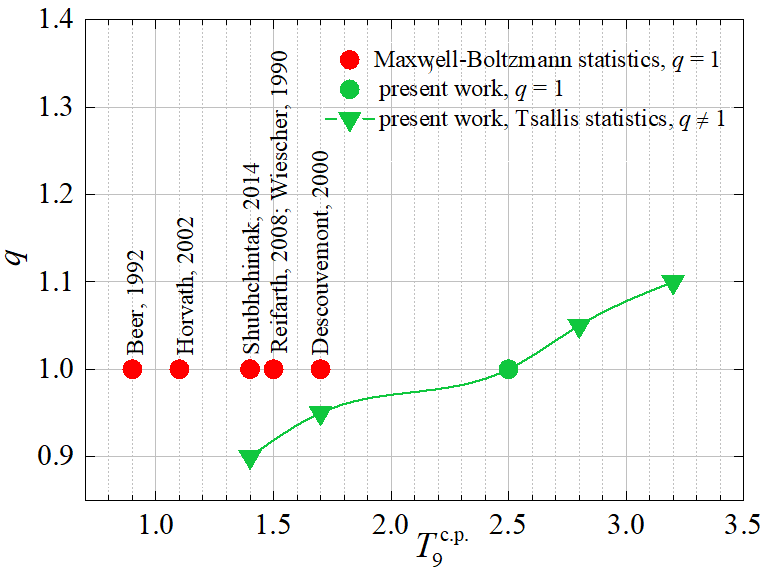}
\caption{$T_9^{c.p}$ temperature relevant to equality of $^{14}\mathrm{C}(n,\gamma_{0+1})^{15}\mathrm{C}$ and $^{14}\mathrm{C}(p,\gamma_{0})^{15}\mathrm{N}$ reaction rates depending on the $q$ parameter. Red points are data from Refs. \cite{Ref19,Ref20,Ref32,Ref34,Ref22,Ref64} 
compared with Kawano et al., 1991 \cite{Ref33} 
$^{14}\mathrm{C}(p,\gamma_{0})^{15}\mathrm{N}$. Green symbols refer to present calculations. }
\label{fig:Table}
\end{figure}


%
%
%
The
$^{14}$C$(n,\gamma)^{15}$C
reaction rate used in the present analysis was obtained in our recent MPCM study
~\cite{Ref24}, employing the same theoretical framework as for the
$^{14}$C$(p,\gamma)^{15}$N calculation. The validated neutron-capture rate is now compared with the newly calculated proton-capture rate to determine the temperature at which $^{14}$C$(p,\gamma)^{15}$N 
becomes the dominant pathway for the production of $^{15}$N. 
Using the calculated cross sections, thermonuclear reaction rates were evaluated for temperatures
$0.01\le T_9\le10$.
Figure~\ref{fig:Table} compares the Maxwell--Boltzmann reaction rates for the competing
$^{14}\mathrm{C}(n,\gamma)^{15}\mathrm{C}$
and
$^{14}\mathrm{C}(p,\gamma)^{15}\mathrm{N}$
reactions.
The neutron-capture channel dominates at low temperatures because of the absence of the Coulomb barrier, whereas the proton-capture rate increases rapidly with temperature and eventually exceeds the neutron-capture rate. Since both reactions are calculated within the same theoretical framework, the obtained transition temperature is free from systematic uncertainties associated with combining reaction rates from different theoretical models.

The temperature at which the
$^{14}\mathrm{C}(n,\gamma)^{15}\mathrm{C}$
and
$^{14}\mathrm{C}(p,\gamma)^{15}\mathrm{N}$
reaction rates become equal is referred to as the
\emph{cross-point temperature},
$T_{9}^{\mathrm{c.p.}}$.
The concept was first introduced by Wiescher
\textit{et al.}~\cite{Ref32}.
Using the parameterized
$^{14}\mathrm{C}(n,\gamma)^{15}\mathrm{C}$
and
$^{14}\mathrm{C}(p,\gamma)^{15}\mathrm{N}$
reaction rates from Ref.~\cite{Ref32},
Kawano \textit{et al.}~\cite{Ref33}
obtained
$T_{9}^{\mathrm{c.p.}}=1.5$.
Later, Ref.~\cite{Ref34}
combined several available
$^{14}\mathrm{C}(n,\gamma)^{15}\mathrm{C}$
reaction rates~\cite{Ref19,Ref20,Ref22,Ref64}
with the
$^{14}\mathrm{C}(p,\gamma)^{15}\mathrm{N}$ reaction rate of Kawano
\textit{et al.}~\cite{Ref33},
yielding
$T_{9}^{\mathrm{c.p.}}=0.9$--1.7,
depending on the adopted neutron-capture rate.

Figure~\ref{fig:Table} compares the transition temperatures,
$T_{9}^{\rm c.p.}$,
reported in previous Maxwell--Boltzmann studies with the present self-consistent MPCM prediction. Earlier determinations yielded
$T_{9}^{\rm c.p.}=0.9$--$1.7$,
whereas the present calculation predicts a significantly higher value,
$T_{9}^{\rm c.p.}=2.5$.
The figure also illustrates the influence of non-extensive Tsallis statistics, showing that the transition temperature shifts from
$T_{9}^{\rm c.p.}=1.4$
for
$q=0.90$
to
$T_{9}^{\rm c.p.}=3.2$
for
$q=1.10$.

The higher transition temperature obtained in the present work results from the self-consistent treatment of both competing reactions within the same theoretical framework. In contrast, previous estimates combined neutron- and proton-capture reaction rates obtained from different theoretical models or evaluations, introducing model-dependent systematic uncertainties into the determination of
$T_{9}^{\rm c.p.}$.
The present comparison eliminates this source of uncertainty and therefore provides a more internally consistent determination of the transition between neutron- and proton-capture dominance.

As summarized in Fig.~\ref{fig:Table}, the present value
$T_{9}^{\rm c.p.}=2.5$
lies well above all previous Maxwell--Boltzmann estimates.
This difference reflects the use of a common theoretical framework for calculating both
$^{14}$C$(n,\gamma)^{15}$C
and
$^{14}$C$(p,\gamma)^{15}$N
reaction rates, thereby avoiding inconsistencies associated with combining independently evaluated rates.

\begin{figure}[b]
\centering
\includegraphics[width=8 cm]{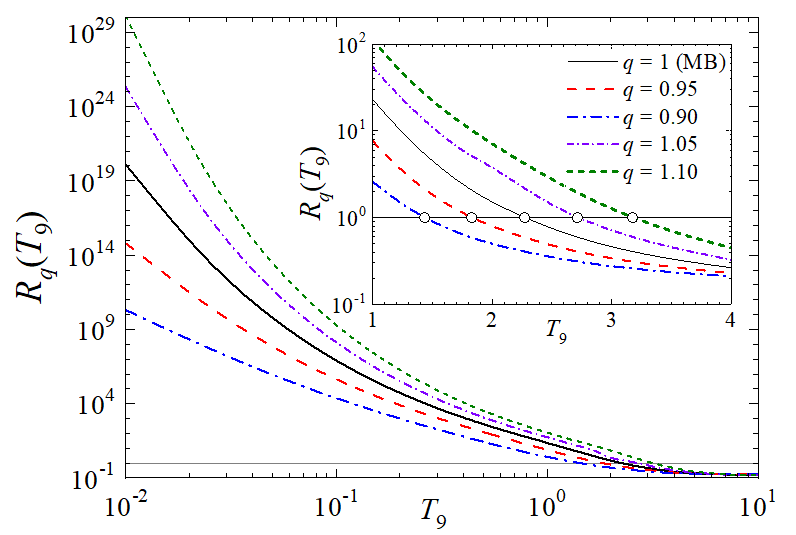}
\caption{
Ratio of the thermonuclear reaction rates
$R_{q}(T_9)$ as a function of temperature for the
$^{14}\mathrm{C}(n,\gamma)^{15}\mathrm{C}$
and
$^{14}\mathrm{C}(p,\gamma)^{15}\mathrm{N}$
reactions.
The solid black curve corresponds to Maxwell--Boltzmann statistics
($q=1$), while the colored curves represent Tsallis statistics for
$q=0.90$, $0.95$, $1.05$, and $1.10$.
The horizontal gray line at
$R_{q}=1$
indicates equal neutron- and proton-capture reaction rates.
Values
$R_{q}(T_9)>1$
correspond to neutron-capture dominance, whereas
$R_{q}(T_9)<1$
indicate proton-capture dominance.
The inset enlarges the temperature interval
$1\leq T_9\leq4$,
where the intersections with
$R_{q}(T_9)=1$
define the cross-point temperature
$T_{9}^{\mathrm{c.p.}}$
for each value of the non-extensive parameter $q$.
}
\label{fig:RateRatioTsallis}
\end{figure}

%
As shown in Fig.~\ref{fig:ReactionRatesComparison}, the proton-capture rate exceeds the neutron-capture rate for
$T_{9}\ge2.5$,
corresponding to
$T_{9}^{\mathrm{c.p.}}=2.5$. This value  is expected to be more reliable because systematic uncertainties associated with combining
reaction rates from different theoretical models are eliminated.

%
%
%

The physical origin of this result is closely related to the different energy windows governing the two reactions.
For the charged-particle reaction
$^{14}\mathrm{C}(p,\gamma)^{15}\mathrm{N}$,
the thermonuclear reaction rate is determined by the effective Gamow window, which arises from the interplay between the Maxwell--Boltzmann energy distribution and the probability of Coulomb-barrier penetration.
At temperatures relevant to AGB stars, this Gamow window lies within the experimentally constrained low-energy region reproduced by the present MPCM calculations, providing a reliable determination of the proton-capture reaction rate.
In contrast, the neutron-capture reaction
$^{14}\mathrm{C}(n,\gamma)^{15}\mathrm{C}$
is not affected by Coulomb-barrier penetration; its reaction rate is governed by the Maxwell--Boltzmann energy distribution together with the energy dependence of the neutron-capture cross section.
The different sensitivities of the two reactions to the underlying energy distribution ultimately determine the cross-point temperature and explain why proton capture becomes dominant only at sufficiently high temperatures.

The Maxwell--Boltzmann analysis establishes the cross-point temperature under thermodynamic equilibrium. To assess the robustness of this result under possible non-equilibrium conditions, we further investigate the competition between the
$^{14}\mathrm{C}(n,\gamma)^{15}\mathrm{C}$
and
$^{14}\mathrm{C}(p,\gamma)^{15}\mathrm{N}$
reactions within the framework of Tsallis statistics. The Tsallis distribution modifies the high-energy tail of the particle-energy distribution relative to the Maxwell--Boltzmann distribution. 
Because only the proton-capture reaction is suppressed by the Coulomb barrier, whereas neutron capture is barrier-free, the two reactions exhibit markedly different sensitivities to the non-extensive parameter $q$.
Consequently, the transition temperature between neutron- and proton-capture dominance,
$T_{9}^{\mathrm{c.p.}}$,
provides a sensitive measure of the influence of non-equilibrium effects on the competition between the two reactions

To display the competition between the two reactions more directly than the
individual reaction rates, we introduce the ratio
\begin{equation}
R_{q}(T_9)
=
\frac{
N_A\left\langle \sigma v \right\rangle_{
^{14}\mathrm{C}(n,\gamma)^{15}\mathrm{C}}
}{
N_A\left\langle \sigma v \right\rangle_{
^{14}\mathrm{C}(p,\gamma)^{15}\mathrm{N}}
}.
\label{eq:RateRatio}
\end{equation}
Here, $q=1$ corresponds to Maxwell--Boltzmann statistics, whereas
$q\neq1$ denotes the Tsallis distribution. The interpretation of
$R_{q}(T_9)$ is immediate:
$R_{q}(T_9)>1
\quad\Longrightarrow\quad
^{14}\mathrm{C}(n,\gamma)^{15}\mathrm{C}$
dominates, 
$R_{q}(T_9)=1
\quad\Longrightarrow\quad$
the neutron- and proton-capture reaction rates are equal,
and
$R_{q}(T_9)<1
\quad\Longrightarrow\quad
^{14}\mathrm{C}(p,\gamma)^{15}\mathrm{N}$
dominates.

Therefore, the intersection of each curve with the horizontal line
$R_{q}(T_9)=1$ directly determines the corresponding cross-point
temperature $T_{9}^{\mathrm{c.p.}}$, at which the dominant production
mechanism changes from neutron capture to proton capture.

The influence of non-extensive Tsallis statistics is illustrated in Fig.~\ref{fig:RateRatioTsallis}. Decreasing the non-extensive parameter
$(q<1)$
enhances the high-energy tail of the particle distribution and increases the proton-capture reaction rate, shifting the transition between neutron- and proton-dominated production of
$^{15}$N toward lower temperatures. In contrast,
$q>1$
suppresses the high-energy tail, delaying the onset of proton-capture dominance. The calculations demonstrate that even relatively small deviations from Maxwell--Boltzmann statistics noticeably modify the competition between the two reaction channels.

The comparison presented here differs fundamentally from previous studies because both competing reactions are treated consistently within the same theoretical framework. This eliminates model-dependent normalization differences between independent reaction-rate evaluations and permits a direct determination of the temperature at which the dominant production mechanism of
$^{15}$N changes from neutron capture to proton capture. The results, therefore, provide improved nuclear-physics input for astrophysical models incorporating both Maxwell--Boltzmann and non-extensive particle distributions.

It is important to emphasize that the cross-point temperature,
$T_{9}^{\mathrm{c.p.}}=2.5$,
determined in the present work corresponds to the purely nuclear-physics condition that the thermonuclear reaction rates are equal,
$N_A\langle\sigma v\rangle_{n\gamma}
=
N_A\langle\sigma v\rangle_{p\gamma}$,
%
which implicitly assumes equal neutron and proton number densities,
$n_n=n_p$.
In an actual stellar environment, however, the reaction rates per unit volume depend on the respective abundances of protons and neutrons. Therefore, the effective competition is governed by the scaled ratio $\frac{n_n}{n_p}R_q(T_9)$. Consequently, in hydrogen-rich environments where
$n_p\gg n_n$, the proton-capture reaction can dominate at temperatures substantially below
$T_{9}^{\mathrm{c.p.}}=2.5$,
whereas in neutron-rich environments the transition shifts to higher temperatures. Thus, the value
$T_{9}^{\mathrm{c.p.}}=2.5$
should be regarded as a fundamental nuclear-physics reference point that can be readily rescaled for specific astrophysical conditions characterized by a given proton-to-neutron density ratio,
$n_p/n_n$.

\textbf{\textit{Conclusions}}. 
%
%
%
%
We have performed the first self-consistent comparative study of the
$^{14}$C$(n,\gamma)^{15}$C
and
$^{14}$C$(p,\gamma)^{15}$N
reactions within the same MPCM. The calculated proton-capture cross section reproduces the available experimental data and yields
$S(0)=4.5(1)$~keV$\cdot \text{b}$. Together with our recent MPCM calculation of the
$^{14}$C$(n,\gamma)^{15}$C
reaction, this provides a consistent determination of the competing thermonuclear reaction rates in both Maxwell--Boltzmann and Tsallis statistics.

The calculated neutron astrophysical
$S_n$-factor and $^{14}$C$(n,\gamma)^{15}$C cross sections agree with the recent
\emph{ab initio}
results of Navrátil \textit{et al.},
supporting the reliability of the neutron-capture reaction rate. Comparison of competing reaction rates predicts a cross-point temperature
$T_{9}^{\mathrm{c.p.}}=2.5$,
significantly higher than previous estimates.  The influence of Tsallis statistics is substantial, shifting the transition temperature from
$T_{9}^{\mathrm{c.p.}}=1.4$
for
$q=0.90$
to
$T_{9}^{\mathrm{c.p.}}=3.2$
for
$q=1.10$.

Finally, we emphasize that the calculated cross-point temperature represents a fundamental nuclear-physics reference corresponding to equal neutron and proton number densities. In specific astrophysical environments, the effective transition temperature can be readily obtained by scaling the competing reaction rates according to the neutron-to-proton abundance ratio.

The present work establishes a reliable and internally consistent framework for determining the transition between competing neutron- and proton-capture reactions and provides improved nuclear-physics input for astrophysical nucleosynthesis calculations under both equilibrium and non-equilibrium conditions.

\section*{Acknowledgments}
This research was funded by the Science Committee of the Ministry of Science and Higher Education of the Republic of Kazakhstan, Grant No. AP23486219.


\begin{thebibliography}{99}
\bibitem{Ref1}
R.A. Malaney, G.J. Mathews, Probing the early universe: A review of primordial nucleosynthesis beyond the standard Big Bang. Phys. Rep. \textbf{229}, 145 (1993).

\bibitem{Ref2}
T. Rauscher,
\textit{Essentials of Nucleosynthesis and Theoretical Nuclear Astrophysics},
IOP Publishing, Bristol, 2020.

\bibitem{Ref3}
C. Iliadis,
\textit{Nuclear Physics of Stars},
2nd ed.,
Wiley-VCH, Weinheim, 2015.

\bibitem{Ref4}
D. Romano, The evolution of CNO elements in galaxies. Astron. Astrophys. Rev. \textbf{30}, 7 (2022).

\bibitem{Ref6}
N. Liu, R. Gallino, S. Cristallo, S. Bisterzo, A. M. Davis, R. Trappitsch, L. R. Nittler, New constraints on the major neutron source in low-mass AGB stars. Astrophys. J. \textbf{865}, 112 (2018).

\bibitem{Ref7}
H.-L. Lu, Y.-B. Li, A.-L. Luo, Z.-Q. Zou, X.-M. Kong, Z.-P. Yi, H. R. A. Jones, J.-C. Liang, S. Li, Review of the search for AGB stars. Front. Astron. Space Sci. \textbf{12} (2025).

\bibitem{Ref19}
H. Beer, M. Wiescher, F. Käppeler,
J. Görres, P.E. Koehler, A measurement of the $^{14}$C($n, \gamma)^{15}$C cross section at a stellar temperature of $kT = 23.3$ keV. Astrophys. J. \textbf{387}, 258 (1992).

\bibitem{Ref20}
A. Horvath, J. Weiner, A. Galonsky, F. Deak, Y. Higurashi, K. Ieki, Y. Iwata, A. Kiss, J.J. Kolata, Z. Seres, Cross section for the astrophysical $^{14}$C(n,$\gamma)^{15}$C reaction via the inverse reaction. 
Astrophys. J. \textbf{570}, 926 (2002).

\bibitem{Ref21}
R. Reifarth, M. Heil, R. Plag, U. Besserer, S. Dababneh, L. Dörr, J. Görres, R.C. Haight, F. Käppeler, A. Mengoni, et al., Stellar neutron capture rates of $^{14}$C.
Nucl. Phys. A \textbf{758}, 787 (2005).

\bibitem{Ref22}
R. Reifarth, M. Heil, C. Forssén, U. Besserer, A. Couture, S. Dababneh, L. Dörr, J. Görres, R.C. Haight, F. Käppeler, et al., The $^{14}$C(n,$\gamma$) cross section between 10 keV and 1 MeV. 
Phys. Rev. C \textbf{77}, 015804 (2008).

\bibitem{Ref23}
T. Nakamura, N. Fukuda, N. Aoi, N. Imai, M. Ishihara, H. Iwasaki, T. Kobayashi, T. Kubo, A. Mengoni, T. Motobayashi, et al., Neutron capture cross section of $^{14}$C of astrophysical interest studied by coulomb breakup of $^{15}$C. Phys. Rev. C \textbf{79}, 035805 (2009).

\bibitem{Ref24}
A.S. Tkachenko, N.A. Burkova,
B.M. Yeleusheva, S.B. Dubovichenko, Estimation of the effect of Tsallis non-extensive statistics on the $^{14}$C(n,$\gamma)^{15}$C reaction rate. Front. Phys. \textbf{13} (2025).

\bibitem{Ref25}
Y. Jiang, Z. He, Y. Luo, W. Xin, J. Chen, X. Li, Y. Shen, B. Guo, G. Li, D. Pang, et al., New determination of the $^{14}$C(n,$\gamma)^{15}$C reaction rate and its astrophysical implications. 
Astrophys. J. \textbf{989}, 231 (2025).

\bibitem{Ref26}
P. Navrátil, S. Quaglioni,
G. Hupin, M. Gennari,
K. Kravvaris, Halo nuclei from an \textit{ab initio} nuclear theory. Particles \textbf{9}(2), 57 (2026).

\bibitem{Ref27}
T. Ma, B. Guo, D. Pang, Z. Li, Y. Su, X. Li, Y. Shen, Y. Wang, Y. Li, J. Su, New determination of astrophysical $^{14}$C(n,$\gamma)^{15}$C reaction rate from the spectroscopic factor of $^{15}$C. Sci. China Phys. Mech. Astron.
\textbf{63}, 212021 (2020).

\bibitem{Ref28}
A. Bhattacharyya, U. Datta,
A. Rahaman, S. Chakraborty, T. Aumann, S. Beceiro-Novo, K. Boretzky, C. Caesar, B.V. Carlson, W.N. Catford, Neutron capture cross sections of light neutron-rich nuclei relevant for r-Ppocess nucleosynthesis. Phys. Rev. C \textbf{104}, 045801 (2021).

\bibitem{Ref49}
N.K. Timofeyuk, D. Baye,
P. Descouvemont, R. Kamouni,
I.J. Thompson, $^{15}$C--$^{15}$F charge symmetry and the $^{14}$C($n$,$\gamma)^{15}$C reaction puzzle.
Phys. Rev. Lett. \textbf{96}, 162501 (2006).

\bibitem{Ref29}
J. Görres, S. Graff,
M. Wiescher, R.E. Azuma, C.A. Barnes, H.W. Becker, T.R. Wang,
Proton capture on $^{14}$C and its astrophysical implications. Nucl. Phys. A \textbf{517}, 329 (1990).

\bibitem{Ref30}
C. Iliadis, R. Longland,
A.E. Champagne, A. Coc,
R. Fitzgerald, Charged-particle thermonuclear reaction rates: II. Tables and graphs of reaction rates and probability density functions. Nucl. Phys. A \textbf{841}, 31 (2010).

\bibitem{Ref31}
C. Iliadis, R. Longland,
K. Setoodehnia, C. Marshall,
P. Mohr, A. Psaltis,  The 2025 evaluation of experimental thermonuclear reaction rates (ETR25).  
Astrophys. J. Suppl. Ser.
\textbf{283}, 17 (2026).

\bibitem{Ref32}
M. Wiescher, J. Görres,
F.-K. Thielemann, Capture reactions on C-14 in nonstandard Big Bang nucleosynthesis.  
Astrophys. J. \textbf{363}, 340 (1990).

\bibitem{Ref33}
L.H. Kawano, W.A. Fowler,
R.W. Kavanagh, R.A. Malaney, Signatures of inhomogeneity in the early universe. Astrophys. J. \textbf{372}, 1 (1991).

\bibitem{Ref34}
Shubhchintak, Neelam,
R. Chatterjee, Capture cross-section and rate of the $^{14}$C(n,$\gamma)^{15}$C reaction from the coulomb dissociation of $^{15}$C. 
Pramana J. Phys. \textbf{83}, 533 (2014).

\bibitem{Ref35}
S.B. Dubovichenko,
\textit{Thermonuclear Processes in Stars and Universe},
2nd English ed.,
Scholar's Press, Saarbrücken, 2015.

\bibitem{RefXX}
S.B. Dubovichenko,
R.Ya. Kezerashvili,
N.A. Burkova,
A.V. Dzhazairov-Kakhramanov,
A.S. Tkachenko, Reanalysis of the $^{13}\mathrm{N}(p,\gamma)^{14}\mathrm{O}$ reaction and its role in the stellar CNO cycle. 
Phys. Rev. C \textbf{102}, 045805 (2020).

\bibitem{RefYY}
S.B. Dubovichenko,
A.S. Tkachenko,
R.Ya. Kezerashvili,
N.A. Burkova,
A.V. Dzhazairov-Kakhramanov, $^{6}\mathrm{Li}(p,\gamma)^{7}\mathrm{Be}$ reaction rate in the light of the new data of the Laboratory for Underground Nuclear Astrophysics. 
Phys. Rev. C \textbf{105}, 065806 (2022).

\bibitem{Ref62}
S. B. Dubovichenko,
A. S. Tkachenko,
R. Ya. Kezerashvili,
N. A. Burkova,
B. M. Yeleusheva, Astrophysical S-Factor and reaction rate for $^{15}$N(p,$\gamma)^{16}$O within the modified potential cluster model.  
Chin. Phys. C \textbf{48}, 044104 (2024).

\bibitem{Ref66}

S. Q. Hou, J. J. He,
A. Parikh, D. Kahl, C. A. Bertulani, T. Kajino, G. J.  Mathews, G. Zhao, Non- extensive statistics to the cosmological lithium problem. Astrophys. J. \textbf{834}, 165 (2017).

\bibitem{Ref61}
Y. J. Li, Z. H. Li, E. T. Li, et al., New determination of the astrophysical $^{13}$C(p,$\gamma)^{14}$N $S(E)$ factors and reaction rates via the $^{13}$C($^7$Li,$^6$He)$^{14}$N reaction.  
Europ. Phys. J. A \textbf{48}, 13 (2012). 

\bibitem{Ref64}
P. Descouvemont, Microscopic cluster study of the C isotopes. Nucl. Phys. A \textbf{675}, 559 (2000).





\end{thebibliography}
\end{document}